\documentclass[floats, aps, showpacs, twocolumn, superscriptaddress, prx,
    reprint, floatfix]{revtex4-1}

\usepackage{graphicx}
\usepackage{amsmath, amssymb, amstext, amsthm, amsfonts}
\usepackage{bbm}
\usepackage[multidot]{grffile}
\usepackage{hyperref}

\newcommand{\HIDDEN}[1]{}
\newcommand\given[1][]{\:#1\vert\:}

\usepackage{lipsum}
\usepackage{color}
\begin{document}

\title{Micro, Meso, Macro: the effect of triangles on communities in networks}

\newcommand{\affilSYD}{
    School of Mathematics and Statistics,
    University of Sydney, 2006 NSW, Australia}

\author{Sophie Wharrie}
\affiliation{\affilSYD}

\author{Lamiae Azizi}
\affiliation{\affilSYD}
  
\author{Eduardo G. Altmann}
\affiliation{\affilSYD}

\date{\today}

\begin{abstract}
Meso-scale structures (communities) are used to understand the macro-scale properties of
complex networks, such as their functionality and formation mechanisms. Micro-scale
structures are known to exist in most complex networks (e.g., large number of triangles or motifs), but they are absent in the simple random-graph
models  considered (e.g., as null models) in  community-detection algorithms.  In this paper we investigate the effect of micro-structures  on the appearance of communities in networks. We find that alone the presence of triangles leads to the appearance of communities even in methods designed to avoid the detection of communities in random networks. 
This shows that communities can emerge spontaneously from simple processes of motiff generation happening at a micro-level. Our results are based on four widely used community-detection approaches (stochastic block model, spectral method, modularity maximization, and the Infomap algorithm) and three different generative network models (triadic closure, generalized configuration model, and random graphs with triangles).
\end{abstract}

\maketitle

\noindent

\section{Introduction}%
A popular approach to understanding the {\it macro-scale} organization of complex networks is to consider their division into {\it meso-scale} communities of nodes. Communities are often defined as groups of nodes that are densely connected within each group but with fewer links between groups, although more general definitions exist~\cite{Fortunato16}.
The problem of detecting communities in networks has received great attention in the last few decades, with many methods and applications across sociology, biology, and computer science ~\cite{Fortunato10,Coscia11}. 

The proposed approaches to identify communities in networks have different partitioning strategies as well as different topological requirements for the definition of a community, resulting in different methods often detecting different communities in the same network.
A fundamental challenge faced by all methods is how to account for (random) fluctuations in the connectivity pattern across nodes.
For instance, a strong limitation of the popular {\it modularity maximization}
approach~\cite{Newman04} is that it detects communities even in Erd\"os-Renyi random
graphs~\cite{Guimera04} (despite accounting for links due to random chance in the definition of modularity~\cite{Newman04,BarabasiBook}). This limitation can be overcome when communities are
inferred from an underlying generative model, such as the stochastic block model
(SBM)\cite{Peixoto18}.
However, the random-graph models used for comparisons in these developments are extremely simple and lack {\it micro-scale} structures known to exist in real networks, such as a positive clustering coefficient or disproportional number of small sub-graphs (motifs).

Another crucial question is how the community detection results are related to the generative process of those networks~\cite{Fortunato10}. The observation of communities in a network are often implicitly associated to the existence of an underlying relationship between the nodes, that influenced the formation of the network and was responsible for the appearance of the community  (e.g., communities in social networks are assumed to reflect some underlying identity between the individuals of a community). This association is far from necessary because complex {\it macro-scale} structures often emerge spontaneously from simple {\it micro-scale} interactions, as observed in multiple examples studied in Complex Systems and Statistical Physics (e.g., collective motion and self-organized criticality).

In this paper, we investigate the effect of triangles on the appearance of communities in networks. This is a further step
in the more general exploration~\cite{Foster11,Bianconi14} between the effect of micro-scale structures -- whose
origin require only local information and are usually more easily explained -- on
meso-scale structures such as the communities found using different algorithms. Our
conjecture is that observed communities in many real networks are an emergent property of 
micro-scale processes. Our paper demonstrates through the analysis of simple models how
this indeed happens. Differently from Refs.~\cite{Foster11,Bianconi14}, we use a variety of community-detection methods, including methods that are robust against the detection of communities in random networks. This is essential to isolate the effect of triangles from the effect of random fluctuations on the formation of communities.

In the next section we describe the different models we consider to generate networks with a tunable clustering coefficient (number of triangles). We then describe the outcome of four different community detection methods on these models, showing the effect of triangles on the communities. We then discuss in further detail our main numerical finding, a phase transition for the number of communities found using an SBM inference-based method. Finally, we summarize our results and discuss their implications. Codes used in our analysis are available in Ref.~\cite{code}.

\section{Generating networks with clustering}\label{sec.models}

In this section we discuss how we generate networks with a tunable amount of triangles. We are interested in simple graphs of $N$ nodes defined by the adjacency matrix $A = \{a_{i,j}\}$, where $a_{i,j}=1$ if there is a link between nodes $i$ and $j$ and $a_{i,j}=0$ otherwise. A triangle exists between nodes $i$, $j$, and $k$ if $a_{i,j}=a_{j,k}=a_{k,i}=1$. The density of triangles in a network can be quantified by the (global) clustering coefficient
\begin{align}\label{eq:globalclustering}
C = \frac{3 \times N_\triangle}{N_3},
\end{align}
where $N_\triangle$ is the total number of triangles in the network, and $N_3 = \sum_{i=1}^N {k_i\choose 2}$ is the number of connected triples (where $k_i=\sum_{j=1}^N a_{i,j}$ is the degree of node $i$).
The clustering coefficient (\ref{eq:globalclustering}) is thus a proxy for the number of
triangles, the quantity we wish to vary. We focus on triangles because of their simplicity,
their simple interpretation within the context of real networks, and the fact that many
real networks with community structure, particularly social networks, also have high
clustering coefficients \cite{Foster11, Newman03}. In fact, the allegoric representation
of triangles in social networks -- ``friends of my friends are also my friends'' -- is also a likely explanation for the process of generating communities (friendship groups).

To support our claims and assess the effect of triangles on the appearance of communities, we generate networks with tunable clustering coefficient (\ref{eq:globalclustering}), then apply four widely used community detection methods to these generated networks. We are interested in models for which we can increase the number of nodes $N$ for a fixed average degree $\langle k \rangle = \frac{1}{N}\sum_{i=1}^N k_i$ (sparse network) and tune the clustering coefficient from 0 to a $C_{max}>0$. 
Considering these constraints we generate networks from three different generative models: (i) the {\bf triadic closure model}~\cite{Bianconi14} in which nodes are added to the network in such a way that new links form triangles with probability $p$; (ii) an extension of the traditional {\bf configuration model} that includes triangles~\cite{Newman09,Miller09}; and (iii) $k$-regular graphs with a fixed number of triangles ({\bf random network with triangles})~\cite{Fischer15}. The triadic closure model is a growth model (nodes are added one by one), whereas the other two generate networks of pre-defined size $N$ with certain constraints. More specifically, the configuration model constrains the (joint) degree distribution $\{t_i, s_i\}$ for the number of triangles and independent edges of each node $i$, while the random network with triangles model imposes the total number of triangles in the network as a hard constraint. Further details on these models can be found in Appendix~\ref{app.models}.

The networks generated by our three models are {\it sparse} networks (i.e. the actual number of links is much smaller than the maximum possible number $N(N-1)/2$), which is a characteristic of most real networks and those generally used in community detection problems. As noted in Ref.~\cite{Bianconi14}, for networks generated by the triadic closure model, groups of nodes become densely connected with triangles as the network grows. In fact for all three of our generative models we expect regions highly concentrated with triangles to create inhomogeneities in the network structure that are detected as communities. 

\section{Communities in networks with clustering}

In order to mimic the usual application of community detection methods in observed networks, we focus on standard community-detection methods and do not use any information about which of the three generative processes described above were used to generate the networks. Differently from previous works~\cite{Fortunato16}, our goal is not to evaluate the performance of the different algorithms, but instead we treat the communities found in each algorithm as given, we analyze them and compare the outcomes of different methods. This approach of using different methods is further motivated by the No Free Lunch Theorem \cite{Peel17}, which stipulates that there is a trade-off on algorithm performance, such that no one method is best for all types of networks.

Our choice of community detection algorithms is guided by the reviews of Fortunato \cite{Fortunato10} and Ghasemian et al. \cite{Ghasemian18}, as well as the practical concern of code availability. This led us to consider four methods, listed in Table \ref{table:algorithms} and summarized in Appendix~\ref{app.methods}. These methods cover the most popular classes of methods discussed in Ref.~\cite{Ghasemian18}.

\begin{figure*}[!bt]
  \includegraphics[width=2\columnwidth]{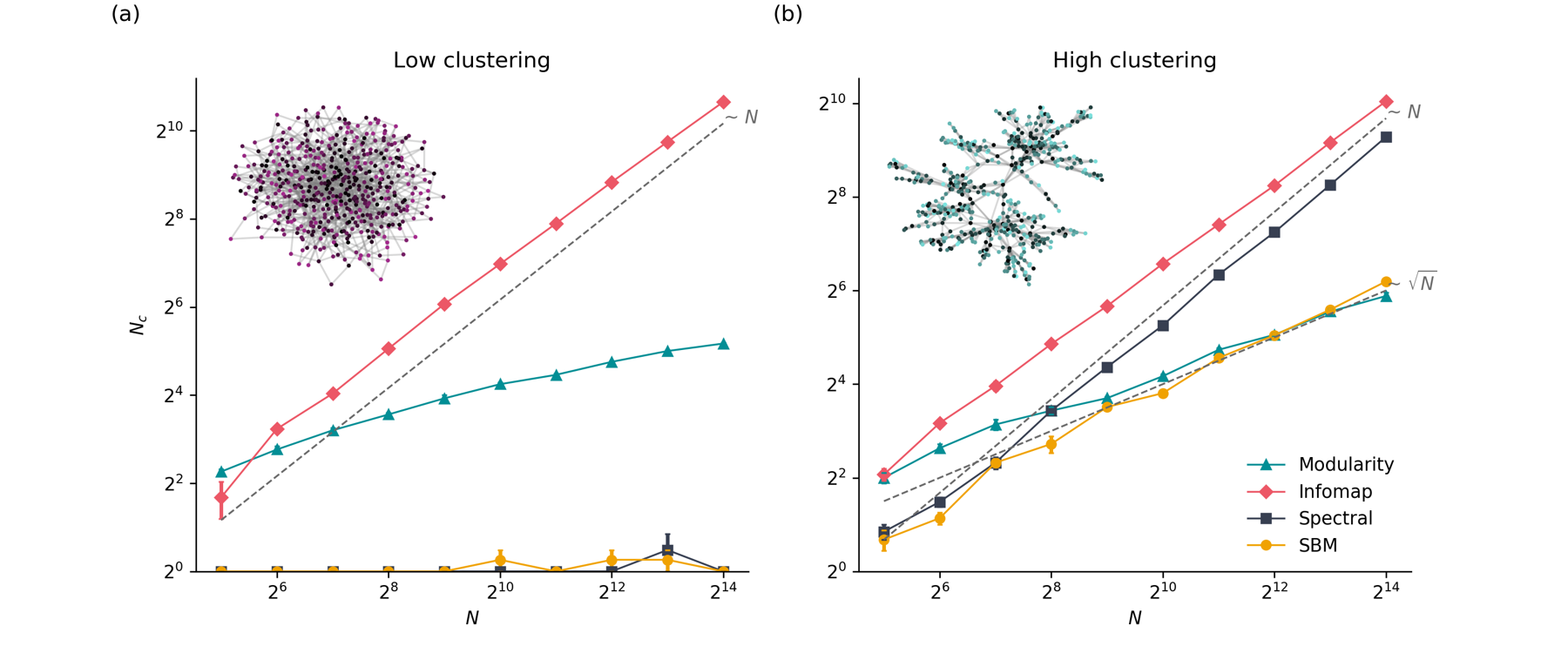}
\caption{Triangles affect the number of communities. The number of detected communities
  $N_c$ as a function of the size of the networks $N$ (number of nodes) for the triadic
  closure model with (a) $p=0$ (low clustering) and (b) $p=1$ (high clustering), for the
  four community detection algorithms. The insets show a sample network obtained in each
  of the cases for $N=500$. Error bars were computed over an ensemble of $5$ different
  networks. The scaling lines show that, in the high clustering case, $N_c$ scales roughly
  linearly with $N$  for the Spectral and Infomap methods, and $N_c$ scales as approximately $\sqrt{N}$ for the Modularity and SBM methods.}
\label{fig.1}
\end{figure*}

\begin{table}[]
\begin{tabular}{l|l}
\textbf{Label} & \textbf{Journal (code) reference}                                             \\ \hline
Modularity maximization   & Clauset et al. \cite{Clauset04} (\cite{Clauset04Web}) \\
Infomap        & Rosvall et al. \cite{Rosvall09} (\cite{Rosvall09Web}) \\
Spectral (Bethe Hessian)      & Saade et al. \cite{Saade14} (\cite{Saade14Web})       \\
SBM      (uninformative priors)      & Peixoto et al. \cite{Peixoto14} (\cite{Peixoto14Web})
\end{tabular}
\caption{Community-detection algorithms used in our analysis (see
  App.~\ref{app.methods} for details).}
\label{table:algorithms}
\end{table}

\begin{figure*}[bt]
  \includegraphics[width=2\columnwidth]{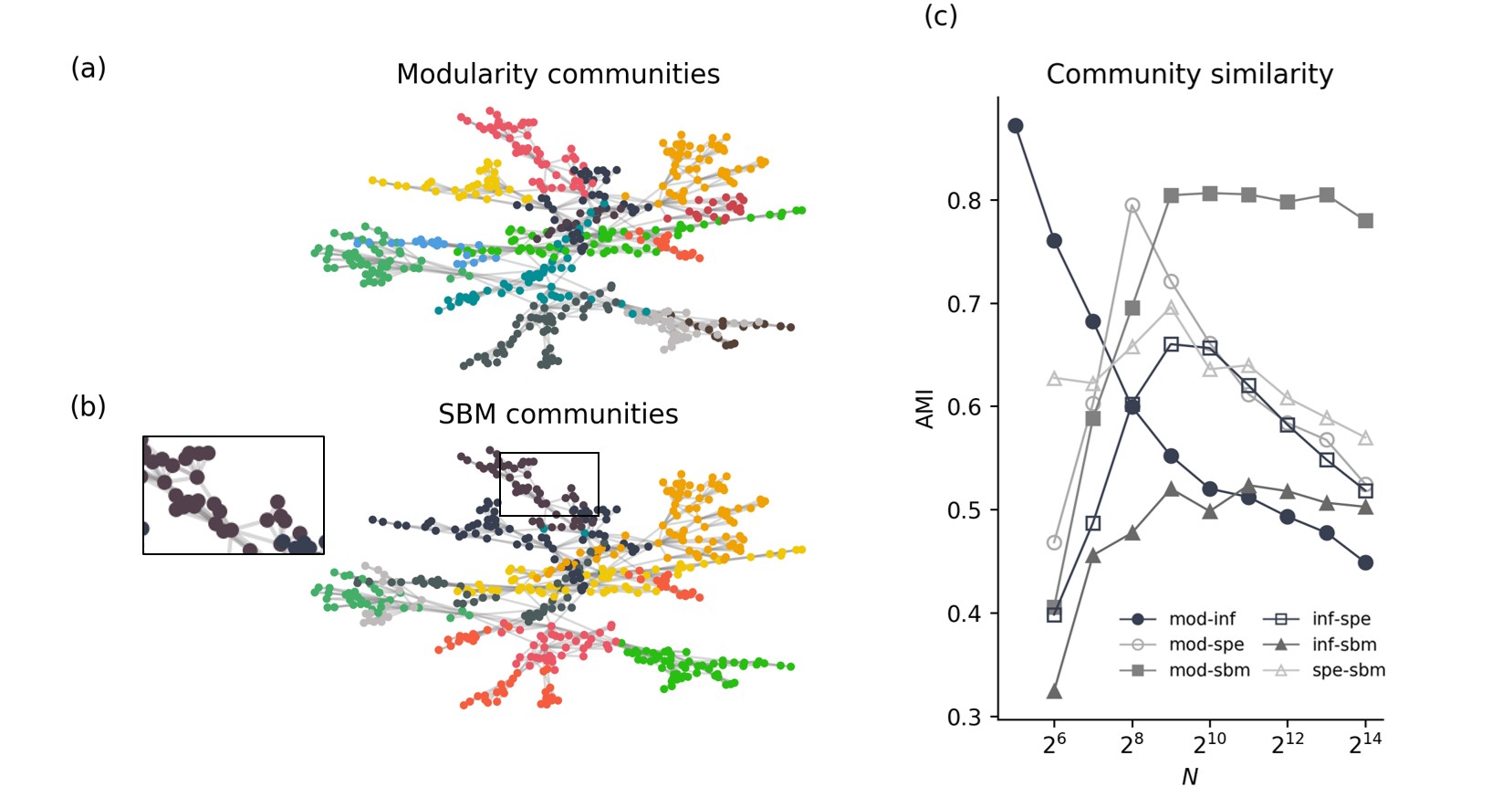} 
\caption{Different methods identify similar communities for the clustered model. (Left) Plots of networks generated by the triadic closure model with $p=1$, with communities obtained using the SBM and modularity maximization methods. The inset emphasizes that triangles are inside communities. (Right) Adjusted Mutual Information between the results of the different methods for various network sizes $N$. The Adjusted Mutual Information~\cite{Vinh10} quantifies the similarity between the communities and varies between $0$ (completely different communities) and $1$ (identical communities), with $0$ indicating the overlap between communities expected by chance.
  }
\label{fig.comparing}
\end{figure*}

As a first assessment of how the presence of triangles leads to the detection of
communities in a network, we investigate the two most contrasting scenarios: networks with
\textit{low clustering} $C \approx 0$ and \textit{high clustering}
$C=C_{max}$. Considering only the triadic closure model as a generative process for now,
we generate networks with $C \approx 0$ (zero probability of closing triangles, $p=0$) and
with $C=C_{max}\approx0.25$ (obtained for $p=1$), then apply the four detection algorithms
described above on the resulting networks. We are interested in investigating the
relationship between the number of communities $N_{c}$ and the network size $N$. Our
results in Fig.~\ref{fig.1} show that for networks with low clustering, only the SBM and
spectral methods result in a single detected community (the whole network). The
low-clustering case is similar to simple random networks and our result thus reflect the
robustness of these two methods to random fluctuations (consistent with previous findings
for the SBM~\cite{Peixoto18}). In contrast, the two other methods (Infomap and Modularity)
detect a number of communities $N_c$ that grows with $N$. As far as high clustering is present in the network, all compared methods behave similarly and $N_c$ grows with $N$, but with different growth scales.  
Interestingly, both the SBM and Modularity methods show similar $N_c$  that grows as $\sqrt{N}$.
This scaling reflects the resolution limit of these methods~\cite{Fortunato07,Peixoto13} and thus corresponds to the maximum number of detectable communities these methods are able to detect in the (sparse) networks generated by the triadic closure model.
Such a growth indicates that for $N\rightarrow \infty$ both the number of communities and
the (average) number of nodes in each community diverge, suggesting that these communities
are indeed {\it meso-}scale structures: more than a fixed number of nodes (the {\it micro} scale) but less than the network as a whole (the {\it macro} scale).  In contrast, when the number of communities grows as $N$ (as our numerical results suggest for the spectral and Infomap cases), the number of nodes in each community remains a constant for $N\rightarrow \infty$.

The similarity of the scaling of the SBM and modularity methods in the case of high
clustering, raises the question whether not only the number but also the communities
themselves are the same. This is investigated in Fig.~\ref{fig.comparing}, which shows
that the communities themselves are very similar to each other. In particular, the
adjusted mutual information~\cite{Vinh10} -- for which 0 indicates no significant
similarity and 1 indicates identical communities -- between the modularity and SBM is
$\approx 0.8$ for large networks ($N>500$). While the comparison between the other methods
leads to less similar results, as expected by the different scalings in the number of
communities, all of them are statistically significant different from $0$ (p-value
$<10^{-3}$).
This happens because the triangles tend to concentrate at the cores of the emerging clusters, with the nodes connected by a triangle belonging predominantly to the same communities and different communities being less connected to each other. This picture is seen in the networks shown in Fig.~\ref{fig.comparing}. This matches the intuition discussed in the previous section,  where we assume that inhomogeneities in the network structure caused by dense clusters of triangles give rise to communities. 

\begin{figure*}[!bt]
  \includegraphics[width=2\columnwidth]{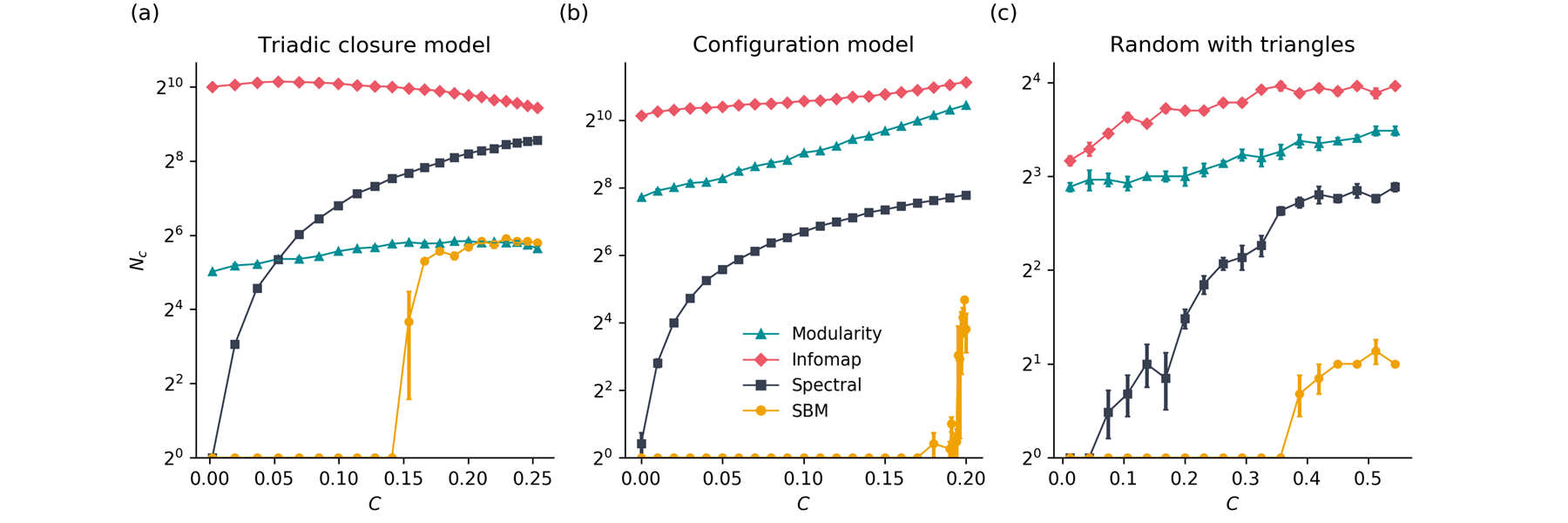}
\caption{Dependence of the number of communities $N_c$ in a network with the clustering coefficient $C$. Each curve corresponds to a different community detection method (see legend and Appendix~\ref{app.methods}). Each panel shows the results for a different network model: (a) triadic closure model with $N=10,000$ and $p\in[0,1]$; (b) the configuration model with $N=10,000$; and (c) the random graph with triangles with $N=80$. The points (error bars) correspond to averages (standard deviation) across 5 different networks (realizations of the network model).}
\label{fig.other}
\end{figure*}

To deepen our investigation on the dependence of communities on
the clustering coefficient $C$, we explore the relationship between the number of
communities~$N_c$ on smooth variations of $C$, from $C=0$ to $C=C_{max}$ for all three generative network models described in section~\ref{sec.models}. The results in Fig.~\ref{fig.other} show that all three cases lead to qualitatively equivalent results, corroborating our claim that the effects we describe here are driven by the existence of triangles, and not by idiosyncratic properties of specific network ensembles. In terms of community detection methods, we find that the two methods that report the existence of communities for $C=0$ (modularity and Infomap) naturally show a smaller dependence on $C$: the  modularity method reports an increasing number of communities as a function of $C$, while Infomap is roughly stable. In contrast, the two methods that do not report the existence of communities for $C=0$ (spectral and SBM) show a starker (and different) dependence on $C$: while for the spectral method the number of communities grows smoothly from $C=0$ on, the SBM method shows a single community for low values of $C$ until a critical (model dependent) value $C=C^*$  for which multiple communities are found. This transition is further investigated in the next section.

\begin{figure}[!bt]
  \includegraphics[width=\columnwidth]{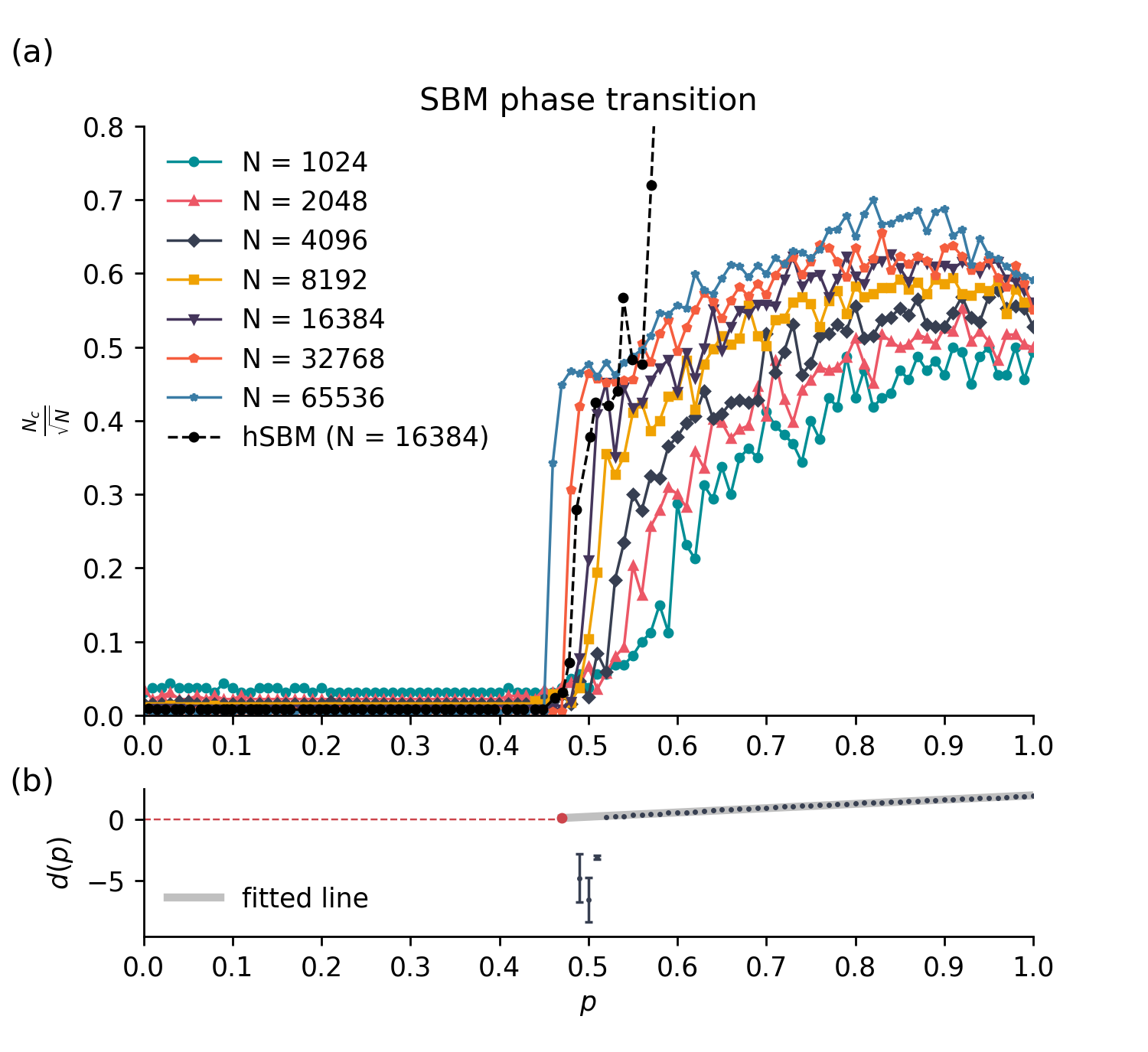}
  \caption{Phase transition in SBM. (a) The scaled number of communities $\frac{N_c}{\sqrt{N}}$ for networks generated by the triadic closure model with varying probability $p$ of closing triangles and different number of nodes $N$ (see legend). As the network size $N$ increases we observe a sharper transition, accumulating around the critical probability $p^* \approx 0.5$ ($C^*\approx 0.15$). The dashed (black) line corresponds to the results of the hierarchical SBM method~\cite{Peixoto14Web,Peixoto18}. (b) The detectability condition $d(p)$ -- Eq.~(\ref{eqn:detectvalue}) -- plotted as a function of $p$, averaged across 5 trials of $N=10,000$ triadic closure networks. Since there is some noise in the data around the critical value $d(p)=0$ (due to the algorithm detecting one large community and a few small communities near the threshold), we fit a line of best fit to the portion of the curve with $d(p) > 0$ and extrapolate to $d(p)=0$. This obtains an estimate (red point) for the critical value $p^*$, which agrees with the  the critical value observed in panel (a). }
\label{fig.sbmtransition}
\end{figure}

\section{Clustering transition in the SBM}

We now explore the abrupt transition observed in Fig.~\ref{fig.other} for the SBM method:
from a single community for $C<C^*$ to a large number of communities ($N_c \sim
\sqrt{N}$) for $C>C^*$. For simplicity, we discuss only the triadic closure model in this
section, since similar results are observed for the other two models. We vary the probability of closing triangles $p$ in the range $p\in[0,1]$, which generates networks with $C\in[0,0.25]$ (see Appendix~\ref{app.models} for the definition of the triadic closure model). In
Fig.~\ref{fig.sbmtransition} we show results for increasing network sizes $N$ -- after
re-scaling the number of communities by $\sqrt{N}$ (the scaling observed in
Fig.~\ref{fig.1}) -- which confirm an increasingly sharp (first order) transition accumulating around
the critical point $p^* \approx 0.5$ ($C^*\approx 0.15$).  Phase transitions in SBMs, and in community
detection methods more generally, have recently been found to describe the transition
between detectable and non-detectable regimes~\cite{Decelle11-1,Krzakala13,
  Peixoto13, Moore17, Zdeborov16,Abbe18}. In this section we further explore the {\it clustering transition} we found in the SBM, establishing its relationship to the {\it detectability transition} investigated previously.

The detectability transition reflects the fact that community detection methods are able
to retrieve partitions planted in the data only if the planted partitions are sufficiently
strong (e.g., many more links within communities than across communities). The previous works
on detectability transitions considered a simplified model ({\it symmetric} SBM~\cite{Abbe18}) for which the $r=1, \ldots, N_c$ blocks (communities) have the same number of nodes $ m = N/N_c$ and the probability of edges between nodes in group $r$ and $s$ is given by $p_I$ if $r=s$ and $p_E$ if $r\ne s$. An assortative community structure is detectable only when $p_I \gg p_E$ (links internal to the communities are much more probable thanks
external links).
More precisely, the $N_c$ symmetric communities used as a planted partition (of a network with average degree $\langle k \rangle$) can be detected only if (see Ref.~\cite{Decelle11-1} and Appendix~\ref{app.sbm})
  \begin{equation}\label{eq.epsilonstar}
    \varepsilon \equiv \frac{p_E}{p_I} < \frac{\sqrt{\langle k \rangle} - 1}{\sqrt{\langle k \rangle} - 1+N_c} \equiv \varepsilon^*,
    \end{equation}
 where $\varepsilon^*$ is a critical value.

While the existence of a regime in which (planted) communities are undetectable has been shown to exist regardless of the community detection method~\cite{Decelle11-1,Krzakala13}, the characteristics of the transition to the detectable regime depend on the specific method. For the main SBM method we use here, which considers a Bayesian setting with uninformative priors for the number of communities $N_c$, the detectability transition has been studied in Ref.~\cite{Peixoto13}. The quantity we are interested here is the number of communities $N_c$, which in this SBM method is determined by minimizing the description length $\Sigma = \mathcal{S} + \mathcal{L}$, where $\mathcal{S}$ is the entropy of the fitted model and $\mathcal{L}$ is the amount of information necessary to describe the model itself. Differently from the fundamental detectability limit discussed in Refs.~\cite{Decelle11-1,Krzakala13}, the transition we observe here corresponds to a transition from $\Sigma > \Sigma \given_{N_c=1}$ (single community $N_c=1$) to  $\Sigma < \Sigma \given_{N_c=1}$ ($N_c>1$).  In Ref.~\cite{Peixoto13} it has been analytically derived (for $E \gg N_c^2$) that this happens when the generated (planted) SBM network satisfies
\begin{align}\label{eqn:detectlimit}
\langle k \rangle > \frac{2 \ln{N_c}}{\mathcal{I}},
\end{align}
where $\mathcal{I} = \frac{1}{2E} \sum_{rs} e_{rs} \ln{\Big(2E \cdot \frac{e_{rs}}{e_r e_s}\Big)}$, $e_{rs}$ is \textit{number of edges} between nodes in blocks $r$ and $s$, $e_r = \sum_s e_{rs}$, and $E$ is the total number of edges in the network.

We now argue that condition~(\ref{eqn:detectlimit}) describes also the clustering
transition we observed. Since our networks were not generated from an SBM planted
partition model, we can only compute $\mathcal{I}$ above using the {\it inferred} SBM
model and check if the inequality is satisfied for these cases. We define
\begin{align}\label{eqn:detectvalue}
d(p) \equiv \langle k \rangle - \frac{2 \ln{N_c}}{\mathcal{I}},
\end{align}
where $\mathcal{I}$ and $N_c$ are computed for the SBM model obtained fitting a network
with clustering $C$. Condition~(\ref{eqn:detectlimit}) corresponds to $d(p)>0$. In the
bottom panel of Fig.~\ref{fig.sbmtransition} we show the results obtained for networks
generated from the triadic closure model. We find that $d(p) > 0$ for all $p>p^*$ and that
$d(p) \rightarrow 0_+$ for $p \rightarrow p^*_+$. This quantitatively connects the phase
transition we found to the detectability transition reported previously. In this
connection, the clustering coefficient (in our network models) plays the role of the
strength of the communities ($p_I \gg p_E$ in the planted SBM models). This is in agreement with the intuition that clustering is related to the probability of intra-community versus inter-community links and further supports our view
that clustering is a driving factor for the appearance of communities in networks.

The results above are specific to the SBM method we use here. We now investigate whether the \textit{clustering transition} appears also for other community-detection methods based on the stochastic block model. In particular, a hierarchical generalization of the SBM method (hSBM) has been proposed and shown to overcome the detectability limit of the method we use (which was responsible for the $N_c \sim \sqrt{N}$ scaling we observe)~\cite{Peixoto18}. Applying this hSBM method to networks generated from the triadic closure model, we observe (dashed line in Fig.~\ref{fig.sbmtransition}) a phase transition at the same critical value $C^*$ (for $C>C^*$ a larger number of communities $N_c \sim N$ is found). This confirms the robustness of the \textit{clustering transition} we found, which appears at the same critical value $p^*$ also in an improved SBM (which overcomes previous detectability limits and is thus more robust against over and under-fitting ~\cite{Peixoto18}).

Further analytical insights on the \textit{clustering transition}, including an estimation of $C^*$, can be obtained considering the symmetric SBM discussed above. This SBM is a good approximation of the SBM we obtain fitting our networks, e.g., for the $C\approx 0.25$ ($p=1$) case we obtain on average (standard deviation) for networks of size $N=10,000$: $n \equiv N/N_c =190 (59)$, $p_I = 0.037 (0.002),$ and $p_E = 0.00002 (0.000001)$. In order to connect the symmetric SBM to the triadic-closure model, we consider that the triangles in the network (responsible for $C>0$) always involve nodes of the same community. This is justified by our results that show that communities are precisely created by these triangles. Using this simplifying assumption, we can compute (see Appendix~\ref{app.pstar}) that $N_c$ symmetric SBM communities will be detected in the triadic closure model only if 
\begin{equation}\label{eq.pstar}
  p>p^* = \frac{1-\varepsilon^*}{\varepsilon^*(N_c-1)+1}=\frac{1}{\sqrt{\langle k\rangle}},
\end{equation}
where $p$ is the probability of a link closing a triangle and $\varepsilon^*$ is defined in Eq.~(\ref{eq.epsilonstar}). This means that, regardless of the number of (symmetric) communities~$N_c$, detectability is expected only if the probability of a link closing a triangle is larger than $1/\sqrt{\langle k \rangle}$. In the cases studied numerically above, $\langle k \rangle = 4$ and thus $p^*=0.5$  (which corresponds to networks with $C \approx 0.151$), in good agreement with the critical values observed numerically in Figs.~\ref{fig.other} and \ref{fig.sbmtransition}.

Finally, we  discuss what distinguishes the clustering transition we found from the
detectability transition reported previously. The main difference is that the networks we
analyze are not generated from an SBM model, as in the usual analysis of detectability transitions. The consequences of this can be best seen by considering what are the properties of the original network that are reproduced by the fitted SBM model. SBM preserves the number of nodes and edges (and thus $\langle k \rangle$), but it fails to reproduce the clustering coefficient $C$ of the original network. For instance, for the maximum clustered network obtained for the triadic closure model ($C\approx 0.25$ obtained for $p=1$), the fitted SBM model shows $C=0.06$ for $N=2000$ (with $N_c=22$ communities found) or $C=0.03$ for $N=10,000$ (with $N_c=54$). The vanishingly small clustering coefficient of the fitted SBM model reflects the fact that within each community the SBM can be viewed as a usual random graph and the number of nodes in each community grows as $n=N/N_c \sim \sqrt{N}$ (because $N_c \sim \sqrt{N}$). In fact,  $C\rightarrow 0$ for $N\rightarrow \infty$ for the symmetric SBM model in the sparse regime considered here (as shown in Appendix~\ref{app.sbm}). It is thus essential to take into account that within each community the SBM does not provide a good description of the networks we are analyzing. More generally, the \textit{clustering transition} we observe is induced by a different process (the clustering coefficient or number of triangles) than the \textit{detectability transition} observed previously for planted (symmetric) SBMs, providing a novel result in the context of community detectability.

\section{Discussion and Conclusion}

In summary, we have investigated the relationship between the clustering coefficient and the number of communities in
complex networks. We found that clustering is a mechanisms for the creation of communities and that networks
that grow following rules at a micro-level (e.g., to close triangles) display an 
emergent appearance of meso-scale structures. The communities we found using four
different community detection algorithms show many similarities (are assortative with
similar partitions), further supporting the idea of the existence of communities in
networks with clustering.  We found communities even in methods that do
not detect communities in (unclustered) random networks, such as the spectral and SBM methods
we use. From the point of view of these methods, our results show that more sophisticated assumptions are needed in case one wants to robustly detect communities that are intrinsically independent from the clustering coefficient.

Our main numerical finding is that the number of communities found by inferring an SBM shows a transition from a single community to multiple communities at a critical clustering coefficient $C^*$. Relating this to previous work on the detectability of communities we find that this phase transition shows similar scalings but also differences, as the fitted SBM models do not reproduce the high clustering found in the original networks. 

Our results demonstrate that communities in networks appear even when the generative process of the network is based on only local information, such as the process of closing triangles. This result is in agreement with the findings of Refs.~\cite{Foster11,Bianconi14}. Our results go beyond these previous findings because we considered a wider class of network models and community detection methods. In fact, the previous works focused on modularity~\cite{Foster11,Bianconi14} and the Infomap method~\cite{Bianconi14} which are now known to be problematic in the detection of the number of communities because of their tendency to find communities even in random networks. Our finding that triangles induce communities also in methods robust to such random fluctuations (SBM and spectral), is thus essential to connect these micro-structures to communities. Interesting future lines of research include the generalization of these findings to other types of motifs and attempts to quantify the role of motifs in the communities found in real networks (e.g., considering community detection methods based on generative models that include clustering).

\begin{acknowledgments}
SW and  EGA were funded by the University of Sydney bridging Grant G199768. We thank M. Gerlach and T. P. Peixoto for insightful remarks and suggestions. 
  \end{acknowledgments}

\appendix

\section{Network models}\label{app.models}

\paragraph{The triadic closure model,} adapted from the basic model of Bianconi et al. \cite{Bianconi14}, takes on three parameters; the number of nodes in the network, $N$, the probability of triadic closure $p$, and the number of links created with each new node, $m$. It is a growth model, initialized with a small connected Erd\"os-R\'enyi random network. At each time step, a new node is added to the network with $m$ links. The first link is attached to a random node of the network. Any subsequent links follow the triadic closure rule: with probability $p$, a link is made to a node \textit{neighboring} a node already connected to the new node, thus closing a triangle. With probability $1-p$, a link is chosen at random from any node not already connected to the new node. This process continues until the network is grown to the desired size $N$. As $p$ increases, the density of triangles in the network increases, implying that the parameter $p$ can be viewed as a tuning mechanism for the clustering coefficient of the network. We use $m=2$ and $p \in [0,1]$.

\paragraph{The configuration model} is extended from the version developed by Newman \cite{Newman09} and Miller \cite{Miller09}. It generates networks with a given joint degree sequence $\{t_i, s_i\}$, with $t_i$ representing the number of triangles in which node $i$ features, and $s_i$ representing the number of additional edges of node $i$ not belonging to these triangles. In our version of the algorithm, the joint degree sequence follows a doubly Poisson distribution
\begin{align}\label{eq:poisson}
p_{st} = e^{-\mu}\frac{\mu^s}{s!}e^{-\nu}\frac{\nu^t}{t!},
\end{align}
where $\nu$ is the average number of triangles per node and $\mu$ is the average number of independent edges per node. Following the method outlined by \cite{Newman09}, this distribution leads to an analytical derivation of the global clustering coefficient
\begin{align}\label{eq:poissonclustering}
C = \frac{2\nu}{2\nu + (\mu + 2\nu)^2},
\end{align}
with an upper bound given by $C_{max} = \frac{1}{1+\langle k \rangle}$, where $\langle k \rangle = \mu + 2\nu$ is the average degree of the network. We can therefore generate networks with a given $C$ by setting $\mu$ and $\nu$ for a fixed value of $\langle k \rangle$ and sampling from (\ref{eq:poisson}). 

\paragraph{Random network with triangles} considers the ensemble of networks with a fixed degree sequence (in our case all nodes have degree $k=4$) and a fixed number of triangles $N_\triangle$. We sample networks from this ensemble using the multicanonical sampling method of Fischer et al. \cite{Fischer15}, which is essential to ensure that the sampled networks are a random choice (over all possibilities in the ensemble) and are also independent from the sampled networks with different $N_\triangle$.

\section{Community-detection methods}\label{app.methods}

\paragraph{The modularity method} identifies the partition of the network that maximizes the Newman-Girvan modularity function \cite{Clauset04}. Our chosen algorithm implements the Newman method for identifying the optimal partition with respect to the modularity (quality) function, which is initialized by assigning each node to its own community. At each step, the algorithm inspects each community pair (connected by at least one link) and joins the pair that achieves the greatest increase in modularity. This process is repeated until no further increase in the modularity  function is achievable, with the resulting partition determining both the number of communities and assignment of nodes into communities.

\paragraph{The Infomap method} follows a similar process (specifically, using the Louvain method), but instead the goal is to minimize the map equation \cite{Rosvall09}. The map equation offers an information-theoretic approach to community detection.

\paragraph{The spectral method} we have chosen utilizes the spectral properties of the Bethe Hessian matrix \cite{Saade14}. The number of communities corresponds to the total number of negative eigenvalues, while the community partition is embedded in the corresponding eigenvectors.

\paragraph{The SBM} method we considered performs community detection by inferring the parameters of a non-hierarchical degree-corrected stochastic block model (SBM) \cite{Peixoto14}. The algorithm we choose uses a Markov chain Monte Carlo technique to infer the SBM parameters that maximize the posterior distribution $P(b|G)$ that an observed network $G$ was generated by a given partition $b$. Importantly, the model selection part of the method can distinguish between statistically significant community structure and randomness, to avoid overfitting the number of communities.

\section{Symmetric SBM model}\label{app.sbm}

The simplified SBM model we consider has $N_c$ identical blocks with fixed block size $n=N/N_c$. The within block probability of links is $p_I$ and the across block probability is $p_E$, identical to all blocks. We are interested in the case of fixed average degree
\begin{equation}\label{eq.k}
  \langle k \rangle = n p_I + (N-n)p_E = \frac{N}{N_c} (p_I + p_E (N_c-1)).
  \end{equation}

\paragraph{Detectability transition.} Following Ref.~\cite{Decelle11-1}, the planted transition is detectable if
$$ |Np_I -Np_E| > N_c \sqrt{\langle k \rangle}$$
Combining this result with Eq.~(\ref{eq.k}) we retrieve Eq.~(\ref{eq.epsilonstar}).

\paragraph{Clustering.} The (average) clustering coefficient is defined as
\begin{align}\label{eqn:sim4}
\widetilde{C} = \frac{1}{N}\sum_{i=1}^N C_i \text{, where }C_i = \frac{2\triangle_i}{k_i(k_i - 1)},
\end{align}
where $k_i$ is the degree of node $i$ and $\triangle_i$ is the number of triangles containing node $i$. We now estimate $\widetilde{C}$ for the symmetric SBM model. First, we approximate $k_i$  by $\langle k \rangle$. Next, consider a triangle containing $i$, as well as two other nodes $j$ and $k$. To determine $\triangle_i$, we consider four cases: either $i$, $j$ and $k$ are in the same block (contributing $p_I^3(n-1)(n-2)\equiv \triangle_a$ triangles), $i$ and $j$ ($i$ and $k$) are in the same block but $k$ ($j$) in a different block (contributing $p_I p_E^2(n-1)(N-n)\equiv \triangle_b$ triangles), $j$ and $k$ are in the same block but $i$ in a different block (contributing $p_E^2 p_I(N-n)(n-1) \equiv \triangle_c$ triangles), or $i$, $j$ and $k$ are all in distinct blocks (contributing $p_E^3(N-n)(N-n-1)$ triangles). Combining this with the simplifying assumption that $C_i$ is the same for all nodes $i$ we obtain 
\begin{equation}\label{eqn:analyticalclustering}
  \widetilde{C} \approx \frac{2(\triangle_a+\triangle_b+\triangle_c)}{\langle k \rangle(\langle k \rangle - 1)}.
  \end{equation}
From Fig.~\ref{fig.1} we know that $n = N/N_c \sim \sqrt{N}$. Since $\langle k \rangle$ is fixed for all $N$ -- sparse network, see Eq.~(\ref{eq.k}) -- it follows that $p_I \sim 1/n \sim 1/\sqrt{N}$ and  $p_E \sim 1/N$. Considering these scalings, we see that $\triangle_{a,b,c} \rightarrow 0$ when $N\rightarrow 0$ and thus $\widetilde{C} \rightarrow 0$ for $N\rightarrow 0$  in Eq. (\ref{eqn:analyticalclustering}). 

\section{Transition in the triadic closure model}\label{app.pstar}
  We assume that a symmetric SBM model with $N_c$ communities is used to describe the network obtained by the triadic closure model defined in Appendix~\ref{app.models}. In the triadic closure  model, nodes have typically $k_i = \langle k \rangle = 2m$ links and links close triangles with a fixed probability $p$. Each link will be internal to the community of the given node either if it closes a triangle (with probability $p$) or (with probability $1-p$) if it is connected by chance to a node of the same community (probability $1/N_c$, assuming the existence of $N_c$ symmetric groups). The total probability of being internal is thus $p+(1-p)/N_c$. The expected number of links internal to the same group of the node is $e_I = 2m (p   + (1-p)/ N_c)$ and the number of links external to the group is $e_E = 2m (1-p) N_c/(N_c-1)$ (such that $e_I+e_E = 2m = \langle k \rangle$). The symmetric SBM probabilities are computed dividing the expected number of links of the node by the number of nodes in each group as $p_I = e_I/n$ and $p_E= e_E / (n(N_c-1))$, thus leading to
  \begin{equation}\label{eq.pe}
    \varepsilon \equiv \frac{p_E}{p_I} =  \frac{1-p}{pN_c+1-p},
  \end{equation}
which is independent of $m$. Introducing Eq.~(\ref{eq.pe}) in Eq.~(\ref{eq.epsilonstar}) we obtain Eq.~(\ref{eq.pstar}).

\end{document}